\documentclass[prd,twocolumn,showpacs,preprintnumbers,superscriptaddress,floatfix,nofootinbib]{revtex4-1}
\usepackage{amssymb}
\usepackage{amsmath,txfonts}
\usepackage{graphicx}
\usepackage{dcolumn}
\usepackage{color}
\usepackage{bm}
\usepackage[subfigure]{graphfig}
\usepackage{makecell}
\usepackage[colorlinks,
            citecolor=blue,
            anchorcolor=green,
            menucolor=orange,
            linkcolor=red,
            filecolor=red,
            runcolor=pink,
            urlcolor=blue,
            frenchlinks=red]{hyperref}

\begin{document}

\title{\boldmath Investigation of pion-induced $f_1(1285)$ production off a
nucleon target within an interpolating Reggeized approach}
\author{Xiao-Yun Wang}
\thanks{xywang@lut.cn}
\affiliation{Department of Physics, Lanzhou University of Technology,
Lanzhou 730050, China}
\author{Jun He}
\thanks{Corresponding author : junhe@impcas.ac.cn}
\affiliation{Department of  Physics and Institute of Theoretical Physics, Nanjing Normal University,
Nanjing, Jiangsu 210097, China}
\begin{abstract}
In this work, the pion-induced $f_1(1285)$ production off a nucleon target
is investigated in an effective Lagrangian approach with an interpolating
Reggeized treatment in a large range of the pion-beam momentum from
threshold up to several tens of GeV. The $s$-channel, $u$-channel, and $t $%
-channel Born terms are included to calculate production cross sections. An
interpolating Reggeized treatment is applied to the  $t$ channel, which is
found to be important to reproduce the behavior of the existent experimental total
cross sections at both low ($\lesssim$ 8 GeV) and high pion-beam momenta ($%
\gtrsim$ 8 GeV). It is found that the $t$-channel contribution is dominant
in the pion-induced $f_1(1285)$ production at low beam momentum and still
dominant at very forward angles at high  momentum. The interpolated Reggeized treatment of the $u$ channel is also discussed. The $u$-channel
contribution is small and negligible at low  momentum,  and it becomes dominant  at backward angles at  momenta higher than 10 GeV. The differential cross sections are predicted
with the model fixed by the fitting  existent experimental data. The
results are helpful to the possible experiments at J-PARC and COMPASS.
\end{abstract}

\maketitle

\section{Introduction}

%light meson production
The light meson is an important topic in  hadron physics. The large
nonperturbative effect in the light meson makes it relatively difficult to
explore its internal structure. Due to the same reason, it is a wonderful
place to study  nonperturbative QCD. Though great progress has been
achieved in the study of  light meson spectroscopy during the last few
decades, the internal structure of  the light meson is still unclear, such as
the debates about the $\sigma (500)$, $a_{1}/f_{1}(980)$, $\kappa $, and $%
f_{1}(1285)$~\cite{Guo:2017jvc}. Hence, many large experimental facilities
will be working in this research area, such as LHCb, BelleII, and the CEBAF
12 GeV. In particular, a new detector GlueX has been installed at CEBAF after the 12 GeV
upgrade, which will focus on the light meson study with electron or photon beams~\cite{Dudek:2012vr}. Like
the light meson photoproduction off the nucleon, the pion-induced light
meson production is also an important way to study the internal structure of
the light meson. This process is accessible at J-PARC~\cite{Kumano:2015gna}
and COMPASS~\cite{Nerling:2012er} with high-energy secondary pion beams,
which provide a good opportunity to study the light meson  combined with
the high-luminosity experiment at CEBAF with an electromagnetic probe.

%f1(1285) and its production
Among the light mesons,  $f_{1}(1285)$ attracts much attention. Its 
internal structure has been studied for many years and is a
long-standing problem. The Patrignani $et~al.$ (PDG) lists  $%
f_{1}(1285)$ as an axial-vector state with quantum number $%
I^{G}(J^{PC})=0^{+}(1^{++})~$\cite{Olive:2016xmw} . It has been suggested as
a dynamically generated state produced from the $K\bar{K}^{\ast }$
interaction in the literature \cite%
{Roca:2005nm,Roca:2005nm,Geng:2015yta,Zhou:2014ila}. In  recent years,
many XYZ particles were observed in the charmed and bottomed sector, such as
$X(3872)$, $Z_{c}(3900)$, $Z_{c}(4025)$, $Z_{c}(10610)$, and $Z_{c}(10650)$~%
\cite{Choi:2003ue,Ablikim:2013mio,Ablikim:2013emm,Belle:2011aa}.  $%
f_{1}(1285)$ and these XYZ particles  are close to the $K\bar{K}^{\ast }/D%
\bar{D}^{\ast }/B\bar{B}^{\ast }$ thresholds, respectively. The similarity in three flavor
sections suggests that these particles are from the corresponding
hadron-hadron interactions, which is supported by explicit calculations in
the one-boson-exchange model~\cite%
{Lu:2016nlp,He:2014nya,He:2015mja,He:2013nwa,Sun:2011uh,Sun:2012zzd}.
In particular,  $f_{1}(1285)$ is the strange partner of the $X(3872)$ as
S-wave hadronic molecular states from the $K\bar{K}^{\ast }$ and $D\bar{D}%
^{\ast }$ interactions, respectively~\cite{Lu:2016nlp}. Compared with the XYZ
particles in charmed and bottomed sectors,  $f_{1}(1285)$ is quite far from
the $K\bar{K}^{\ast }$ interaction. Hence, more investigation of  $%
f_{1}(1285)$ in different production processes may provide more helpful
information to confirm the molecular state interpretation of  $f_{1}(1285)
$. Recently, the $f_{1}(1285)$ meson was studied at CLAS in 
photoproduction from a proton target, and its decay pattern was extracted
from high-precision data~\cite{Dickson:2016gwc}. A nucleon resonance of a
mass of about 2300 MeV was suggested in the analyses~\cite%
{Dickson:2016gwc,Wang:2017hug}. However, a calculation with an interpolated
Reggeized treatment suggests that the experimental cross section can be well
reproduced without nucleon resonance included~\cite{Wang:2017plf}. To check
different models, an experimental study of  pion-induced production
process will be helpful.

%pion-induced f1(1285) production
Until now, there only exist some old experimental data and no explicit
theoretical study of those data can be found in the literature to our knowledge~%
\cite%
{Dahl:1967pg,Corden:1978cz,Dionisi:1980hi,Bityukov:1983cw,Bityukov:1987bj}.
Furthermore, it is promising to launch new measurements of the pion-induced $%
f_{1}(1285)$ at J-PARC and COMPASS. Hence, it is interesting to analyze the
pion-induced $f_{1}(1285)$ production based on the old data in an effective
Lagrangian approach to provide helpful predictions for the future
experiment. Because the exisiting data scatter from near threshold to serveral
tens of GeV, we will introduce the interpolating Reggeized treatment in the $t
$ channel as in the $f_{1}(1285)$ photoproduction to reproduce the data at
both low and high beam momentum~\cite{Nam:2010au}. The $t$ channel and $u$
channel usually correspond to the enhancement at forward and backward
angles, respectively~\cite{He:2013ksa,He:2014gga}. The only existing data of
the differential cross section are at very forward angles~\cite{Corden:1978cz}, which can be used
to determine the $t$-channel contribution. From the previous studies, the $u$-channel contributions will become more important at higher beam momentum~\cite{He:2013ksa,He:2014gga}. The $u$ channel's
contribution was found to be essential to interpret the behavior of the
differential cross section of photoproduction~\cite{Wang:2017plf}. Hence, in
this work, we will consider the  $u$ channel as well as $t$ and $s$ channels to calculate the
behavior of the pion-induced $f_1(1285)$ production in a large range of the beam momentum. It can be
expected that the Born $s$ channel is negligible. Since the experimental
data are very crude and  information about the coupling constant is lacking,
the $s$-channel nucleon resonance is not included  in the current work as in the $f_{1}(1285)$ photoproduction~\cite{Wang:2017plf} to  keep model  simplified.

This paper is organized as follows. After the Introduction, we present the
formalism including Lagrangians and amplitudes of pion-induced $f_1(1285)$
production in Sec. II. The numerical results of cross sections are
presented in Sec. III and  compared with the existing data. Finally, the
paper{\ ends} with a summary.

\section{Formalism}

\subsection{Lagrangians}

\begin{figure}[tbph]
\begin{center}
\includegraphics[scale=0.52]{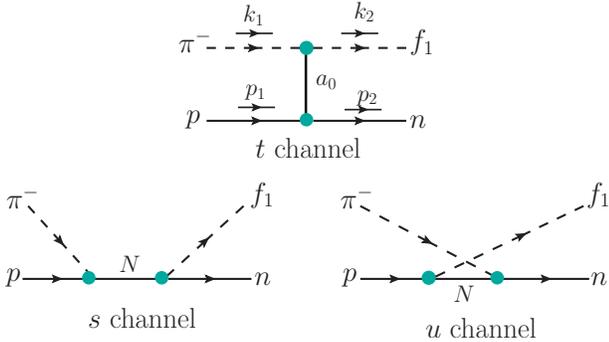}
\end{center}
\caption{Feynman diagrams for the $\protect\pi^{-}p\rightarrow f_{1}(1285)n$
reaction.}
\label{Fig: Feynman}
\end{figure}

The basic tree-level Feynman diagrams for the $\pi ^{-}p\rightarrow
f_{1}(1285)n$ reaction are depicted in Fig.~\ref{Fig: Feynman}. These
include $t$-channel $a_{0}(980)$ ($\equiv a_{0}$) exchanges $s$ and $u$ channels with
intermediate nucleon. As shown by  PDG~\cite{Olive:2016xmw} , the main
two-body decay of  $f_1(1285)$ ($\equiv f_1$) is the $a_0\pi$ channel.
Hence, only the $a_0$ exchange is included in the $t$
channel.

For the $t$-channel $a_{0}$ exchange, one needs the following Lagrangians \cite%
{Liu:2008qx,Penner:2002ma,Colangelo:2010te}%
\begin{eqnarray}
\mathcal{L}_{a_{0}NN} &=&g_{a_{0}NN}\bar{N}(\bm{\tau}\cdot \bm{a}_{0}){N}, \\
\mathcal{L}_{f_{1}a_{0}\pi } &=&-g_{f_{1} {a}_{0}\pi }f_{1}^{\mu }\bm{a}%
_{0}\cdot\partial _{\mu }{\bm \pi},
\end{eqnarray}%
where ${N}$, ${f_{1}}$, $a_{0}$, and $\pi $ are the nucleon, $f_{1}(1285)$, $%
a_{0}(980)$ and $\pi $ meson fields, respectively. The coupling constant $%
g_{f_{1}a_{0}\pi }$ is determined from the decay width%
\begin{equation}
\Gamma _{f_{1}a_{0}\pi }=g_{f_{1}a_{0}\pi }^{2}\frac{%
(m_{f_{1}}^{2}-m_{a_{0}}^{2}+m_{\pi }^{2})^{2}-4m_{f_{1}}^{2}m_{\pi }^{2}}{%
24\pi m_{f_{1}}^{4}}\left\vert \bm{p}_{\pi }^{~\mathrm{c.m.}}\right\vert ,
\end{equation}%
where $\bm{p}_{\pi }^{~\mathrm{c.m.}}$ is the three-momentum of the pion in the
rest frame of the $f_{1}$ meson. By taking the value at PDG as $\Gamma
_{f_{1}\rightarrow a_{0}\pi }\simeq 8.71$ MeV~\cite{Olive:2016xmw}, one gets
a value of the coupling constant $g_{f_{1}a_{0}\pi }\simeq 4.53$. The
coupling constant $g_{a_0NN}$ was not well determined in the literature~\cite%
{Liu:2008qx,Penner:2002ma}. In the current work, we will take  $g_{a_0NN}$
as a free parameter. For the $t$-channel $a_0$ meson exchange, the general form
factors $F_{f_{1}a_0\pi }=(\Lambda _{t}^{2}-m_{a_0}^{2})/(\Lambda
_{t}^{2}-q_{a_0}^{2})$ and $F_{a_0NN}=(\Lambda
_{t}^{2}-m_{a_0}^{2})/(\Lambda _{t}^{2}-q_{a_0}^{2})$ are taken into account
in this work and the cutoffs are taken as the same one for simplification.
Here, $q_{a_0}$ and $m_{a_0}$ are the four-momentum and mass of the exchanged $%
a_0$ meson, respectively.

To calculate the amplitude of the $s$-channel nucleon exchange, we need
relevant Lagrangians. For the $\pi NN$ interaction vertex we take the
effective pseudoscalar coupling \cite{Tsushima:1998jz}%
\begin{equation}
\mathcal{L}_{\pi NN}=-ig_{\pi NN}\bar{N}\gamma _{5}\bm{\tau}\cdot \bm{\pi}{N}%
\text{ },
\end{equation}%
where $\bm{\tau}$ is the Pauli matrix, and $g_{\pi NN}^{2}/4\pi =12.96$ is
adopted \cite{Lin:1999ve,Baru:2011bw}.

The Lagrangian of the $f_{1}NN$ coupling reads ~\cite{Domokos:2009cq},%
\begin{equation}
\mathcal{L}_{f_{1}NN}=g_{f_{1}NN}\bar{N}\left( {f_{1}}^{\mu }-i\frac{\kappa
_{f_{1}}}{2m_{N}}\gamma ^{\nu }\partial _{\nu } {f_{1}}^{\mu }\right) \gamma
_{\mu }\gamma ^{5}{N}+\text{H.c.},
\end{equation}%
where $g_{f_{1}NN}=2.5$ will be taken as discussed in Ref. \cite%
{Birkel:1995ct}. Since the value of $\kappa _{f_{1}}$ was determined by
fitting the CLAS data in our previous work \cite{Wang:2017plf},  $\kappa
_{f_{1}}=1.94$ is adopted in this paper. For the $s$ and $u$ channels with
intermediate nucleons, we adopt the{\ general form factor to describe the
size of the hadrons \cite{Kochelev:2009xz},}%
\begin{equation}
F_{s/u}(q_{N})=\frac{\Lambda _{s/u}^{4}}{\Lambda
_{s/u}^{4}+(q_{N}^{2}-m_{N}^{2})^{2}}~,
\end{equation}%
where $q_{N}$ and $m_{N}$ are the four-momentum and mass of the exchanged
nucleon, respectively. Since the $s$-wave contribution is usually very
small, we take $\Lambda _{s}=\Lambda _{u}$. The values of cutoffs $\Lambda
_{s}$, $\Lambda _{u}$ and $\Lambda _{t}$ will be{\ determined by fitting
experimental data}.

\subsection{Amplitude for $\protect\pi ^{-}p\rightarrow f_{1}(1285)n$
reaction}

The scattering amplitude of the $\pi ^{-}p\rightarrow f_{1}(1285)n$ process
can be written in a general form of%
\begin{equation}
-i\mathcal{M}_{i}=\epsilon _{f_{1}}^{\mu \ast }(k_{2})\bar{u}(p_{2})\mathcal{%
A}_{i,\mu }u(p_{1}),
\end{equation}%
where $\epsilon _{f_{1}}^{\mu \ast }$ is the polarization vector of the $f_{1}$
meson and $\bar{u}$ or $u$ is the Dirac spinor of the nucleon.

The reduced amplitudes $\mathcal{A}_{i,\mu }$ for the $s$-, $t$-, and $u$%
-channel contributions read
\begin{eqnarray}
\mathcal{A}_{s,\mu }^{(N)} &=&-\sqrt{2}g_{\pi
NN}g_{f_{1}NN}F_{s}(q_{N})\left( 1-i\frac{\kappa _{f_{1}}}{2m_{N}}%
\rlap{$\slash$}k_{2}\right)   \notag \\
&\cdot &\gamma _{\mu }\gamma ^{5}\frac{(\rlap{$\slash$}q_{N}+m_{N})}{%
s-m_{N}^{2}}\gamma _{5}, \\
\mathcal{A}_{t,\mu }^{(a_{0})} &=&i\sqrt{2}g_{a_{0}NN}g_{f_{1}a_{0}\pi
}F_{t}(q_{V})\frac{1}{t-m_{a_{0}}^{2}}k_{1\mu },  \label{Texchange} \\
\mathcal{A}_{u,\mu }^{(N)} &=&-\sqrt{2}g_{\pi
NN}g_{f_{1}NN}F_{u}(q_{N})\gamma _{5}\frac{(\rlap{$\slash$}q_{N}+m_{N})}{%
u-m_{N}^{2}}  \notag \\
&\cdot &\left( 1-i\frac{\kappa _{f_{1}}}{2m_{N}}\rlap{$\slash$}k_{2}\right)
\gamma _{\mu }\gamma ^{5},  \label{AmpT2}
\end{eqnarray}%
where $s=(k_{1}+p_{1})^{2}$, $t=(k_{1}-k_{2})^{2}$ and $u=(p_{2}-k_{1})^{2}$
are the Mandelstam variables.

\subsection{Interpolating Reggeized $t$ channel}

In this work, we will consider a large beam-momentum range from threshold to
several tens of GeV. To describe the behavior of the hadron production at
high momentum, the Reggeized treatment should be introduced to the $t$
channel\cite%
{He:2010ii,Galata:2011bi,Haberzettl:2015exa,Wang:2015hfm,Wan:2015gsl}. The Reggeized treatment for $t$-channel meson exchange consists of replacing
the product of the form factor in Eq.~(\ref{Texchange}) as
\begin{equation}
F_{t}(t)\rightarrow \mathcal{F}_{t}(t)=(\frac{s}{s_{scale}}%
)^{\alpha _{a_{0}}(t)}\frac{\pi \alpha _{a_{0}}^{\prime }(t-m_{a_{0}}^{2})}{%
\Gamma \lbrack 1+\alpha _{a_{0}}(t)]\sin [\pi \alpha _{a_{0}}(t)]}.
\end{equation}%
The scale factor $s_{scale}$ is fixed at 1 GeV. In addition, the Regge
trajectories $\alpha _{a_{0}}(t)$ read as
$\alpha _{a_{0}}(t)=-0.5~{\rm GeV}^2+0.6t$~\cite{Kochelev:2009xz,Galata:2011bi}.

To describe the behavior of the cross sections at both low and high
beam momentum, an interpolating Reggeized treatment will be adopted to
interpolate the Regge case smoothly to the Feynman case, which has been
successfully to applied to several photoproduction processes\cite%
{Nam:2010au,Haberzettl:2015exa,Wang:2015hfm,He:2012ud,He:2013ksa,He:2014gga}%
. The interpolated Reggeized form factor can then be written as%
\begin{equation}
F_{t}\rightarrow \mathcal{F}_{R,t}=\mathcal{F}_{t}R\left( s,t\right) +F_{t}%
\left[ 1-R\left( s,t\right) \right],\label{Eq: interpolating}
\end{equation}%
where $R\left( s,t\right) =$ $R_{s}\left( s\right) R_{t}\left( t\right) $,
with%
\begin{equation}
R_{s}\left( s\right) =\frac{1}{1+e^{-(s-s_{R})/s_{0}}},\ \ R_{t}\left(
t\right) =\frac{1}{1+e^{-(t+t_{R})/t_{0}}},
\end{equation}%
where $s_{R}$ and $t_{R}$ are the centroid values for the transition
from non-Regge to Regge regimes while $s_{0}$ and $t_{0}$ describe the
respective widths of the transition regions. The four parameters  will be fitted to the experimental data.
The Feynman-type $u$ channel will be adopted first in the fitting procedure, and the Rggeized treatment of the $u$channel will be discussed in Sec.~\ref{Regu}

\section{Numerical results}

With the preparation in the above section, the differential cross section of
the $\pi ^{-}p\rightarrow f_{1}(1285)n$ reaction will be calculated and
compared with the experimental data~\cite%
{Dahl:1967pg,Corden:1978cz,Dionisi:1980hi,Bityukov:1983cw,Bityukov:1987bj}.
The differential cross section in the center-of-mass (c.m.) frame is written
as
\begin{equation}
\frac{d\sigma }{d\cos \theta }=\frac{1}{32\pi s}\frac{\left\vert \vec{k}%
_{2}^{{~\mathrm{c.m.}}}\right\vert }{\left\vert \vec{k}_{1}^{{~\mathrm{c.m.}}%
}\right\vert }\left( \frac{1}{2}\sum\limits_{\lambda }\left\vert \mathcal{M}%
\right\vert ^{2}\right),
\end{equation}%
where $s=(k_{1}+p_{1})^{2}$, and $\theta $ denotes the angle of the outgoing
$f_{1}(1285)$ meson relative to the $\pi $ beam direction in the c.m. frame. $%
\vec{k}_{1}^{{~\mathrm{c.m.}}}$ and $\vec{k}_{2}^{{~\mathrm{c.m.}}}$ are the
three-momenta of the  initial $\pi $ beam and final $f_{1}(1285)$, respectively.
The experimental data for the $\pi ^{-}p\rightarrow f_{1}(1285)n$ reaction
will be fitted with the help of the \textsc{minuit} code in  \textsc{%
cernlib}.

\subsection{\protect\boldmath$t$ distribution for $\protect\pi%
^{-}p\rightarrow f_{1}(1285)n$ reaction}

The interpolating Reggeized treatment is adopted to reproduce the cross
section in a beam-momentum region from threshold to several tens of GeV
considered in the current work. However, four additional free parameters
will be introduced in such treatment. If we recall that at higher beam momenta,
the Reggeized $t$-channel contribution is dominant, the two parameters $t_R$
ad $t_0$ can be determined with the $t$ distribution at a certain beam momentum.
Fortunately, there exist experimental data of the $t$ distribution at
beam momentum in the  laboratory frame $P_{Lab}=12-15$ GeV~\cite{Corden:1978cz}. Hence, we first study $%
t$ distribution and  determine  $t_R$ and $t_0$ before making a full
fitting of all the data points that we collected.

$t_R$ and $t_0$ can be determined by the $t$ distribution  because
 other parameters   only affect  the $s$ dependence at high beam momenta.
Because the experimental data in Ref.~\cite{Corden:1978cz} are at a very high
beam momentum, one can safely assume that  $R_{s}\left( s\right) \approx 1$.
Hence, one minimizes the $\chi ^{2}$ per degree of freedom ($d.o.f.$) for the
total cross section and the $t$ distribution of the experimental data at $%
P_{Lab}=12-15$ GeV by fitting parameters, which include two parameters
for the Regge trajectory $t_{0}$ and $t_{R}$. In Ref~~\cite{Corden:1978cz},
the $t$ distribution is given by the event not the differential cross
section, so a scale parameter should be introduced, which can be related to
the coupling constant $g_{a_{0}NN}$ with the total cross section which was
given in the same Ref.~\cite{Corden:1978cz} (the total cross section is
obtained only by continuation of the $t$-channel contribution at very forward
angles). The cutoff $\Lambda_t$ is also involved through Eq.~(\ref{Eq: interpolating}). Hence, in the calculation we have four parameters as listed in
Table~\ref{tab:fit}.

\renewcommand\tabcolsep{0.3cm} \renewcommand{\arraystretch}{1.5}
\begin{table}[h]
\caption{Fitted values of free parameters by fitting the $t$ distribution in
Ref.~\protect\cite{Corden:1978cz} with a reduced value $\protect\chi %
^{2}/d.o.f.=0.89$.}
\label{tab:fit}%
\begin{tabular}{|c|c|c|c|}
\hline\hline
$\Lambda _{t}$ (GeV) & $g_{a_{0}NN}$ & $t_{0}$ (GeV$^{2}$) & $t_{R}$ (GeV$%
^{2}$) \\ \hline
$1.26\pm 0.05$ & $28.27\pm 2.49$ & $0.41\pm 0.16$ & $1.90\pm 1.62$ \\ \hline
\end{tabular}%
\end{table}

\begin{figure}[h]
\centering
\includegraphics[scale=0.32]{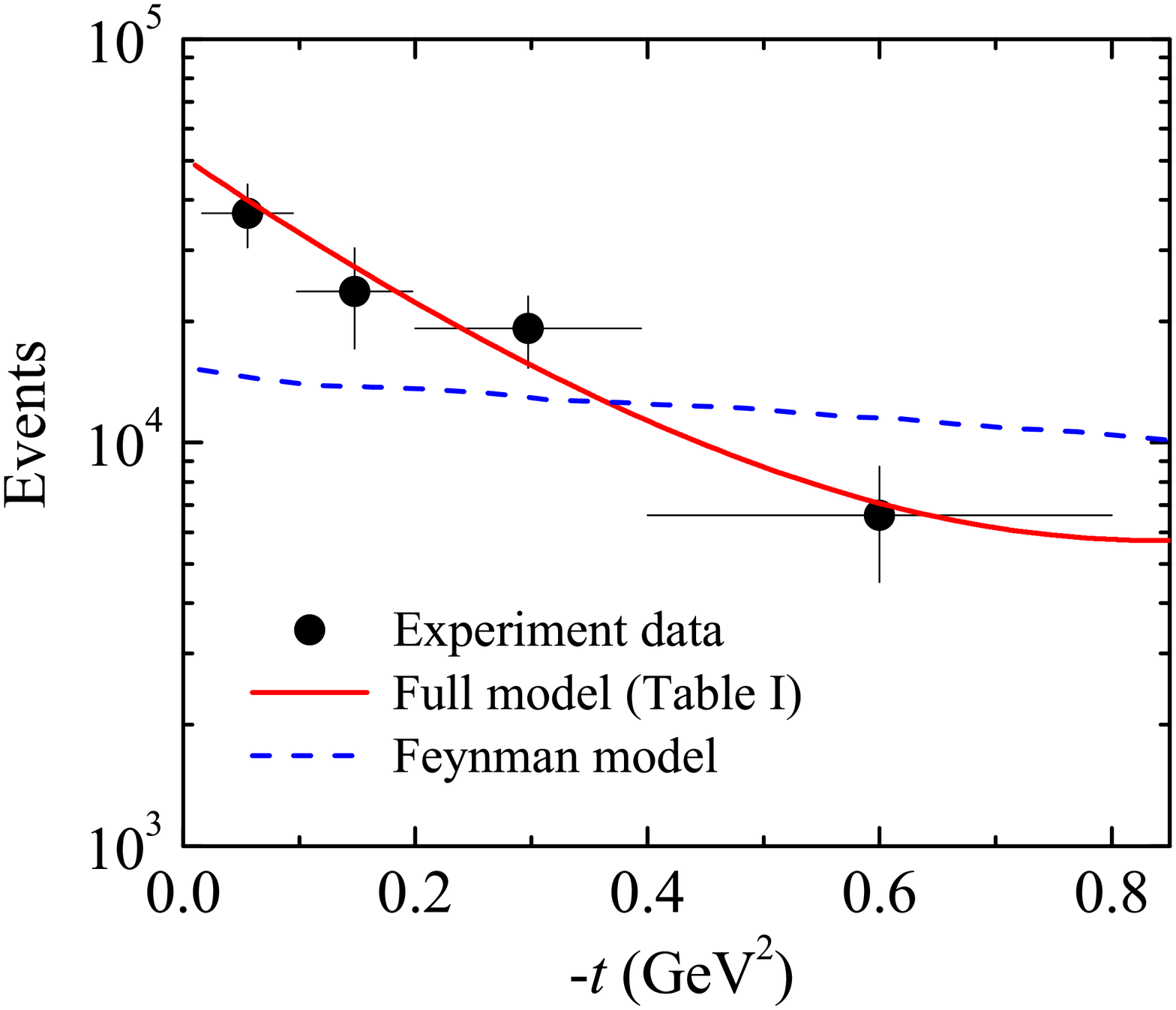}
\caption{The $t$-distribution for the reaction $\protect\pi ^{-}p\rightarrow
f_{1}(1285)n.$ The data are from Ref.~\protect\cite{Corden:1978cz}. The full
(red) and dashed (blue) lines are for the full model and Feynman model,
respectively. }
\label{fig:data}
\end{figure}

The fitted values of the free parameters are listed in Table \ref{tab:fit},
with a reduced value of $\chi ^{2}/d.o.f.=0.89$. The best-fitted results are
presented in Fig.~\ref{fig:data}. It is found that the experimental data of the $%
t$ distribution for the $\pi ^{-}p\rightarrow f_{1}(1285)n$ reaction are
well reproduced in our model. Here, we also present the best-fitted results
with a pure Feynman model. It confirms that at high beam momentum, the results
with a Feynman model deviate from the experimental data obviously, and the
Reggeized treatment is essential to reproduce the $t$-slope.

\begin{figure}[h]
\centering
\includegraphics[scale=0.34]{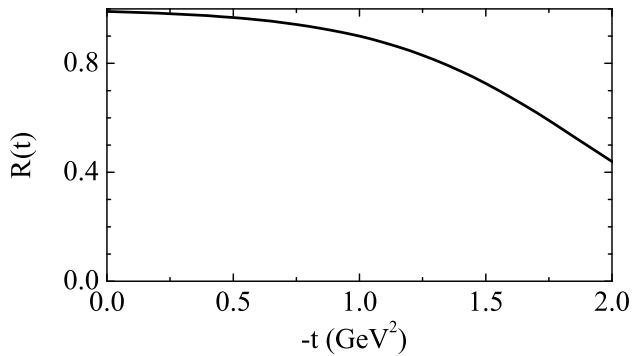}
\caption{Interpolating switching function $R_{t}\left( t\right) $ with the
values of parameters in Table \protect\ref{tab:fit}.}
\label{fig:Rt}
\end{figure}

To show the effect of the interpolating switching function $R_{t}\left(
t\right) $ more clearly, in Fig. \ref{fig:Rt} we present the results with the values of
parameters in Table \ref{tab:fit}. One can see that 
$R_{t}\left( t\right) $ is close to 1 at a small value of $-t$, which
indicates that the contribution of pure Reggeized treatment plays a dominant role
at high beam momentum in the $\pi ^{-}p\rightarrow f_{1}(1285)n$ reaction.

Now we would like to give some discussions about the above results. At low
beam momentum, the $R_{s}(s)$ is very small, which leads to a very small $R(s,t)$%
. Hence, the effect of $R_{t}(t)$ should be small at low beam momentum and becomes
more important at high momentum where $R(s,t)\rightarrow R_{t}(t)$. At high
beam momentum, the $t$-channel contribution is usually dominant at very forward
angles. At medium and backward angles the $u$-channel contribution becomes
more important. The $t_{R}$ of 1.9 GeV$^{2}$ in the $\pi ^{-}p\rightarrow
f_{1}(1285)n$ reaction suggests that at very forward angles the $R(t)$ is
close to 1. Considering  that only a few available data points exist, we will assume
 $R(t)=1$ in the following calculation to reduce the number of free
parameters. It is reasonable because only the results at a medium angle will
be slightly affected where the differential cross section is usually very
small.

\subsection{Cross section of $\protect\pi ^{-}p\rightarrow f_{1}(1285)n$
reaction}

\begin{figure}[h!]
\begin{center}
\includegraphics[scale=0.305]{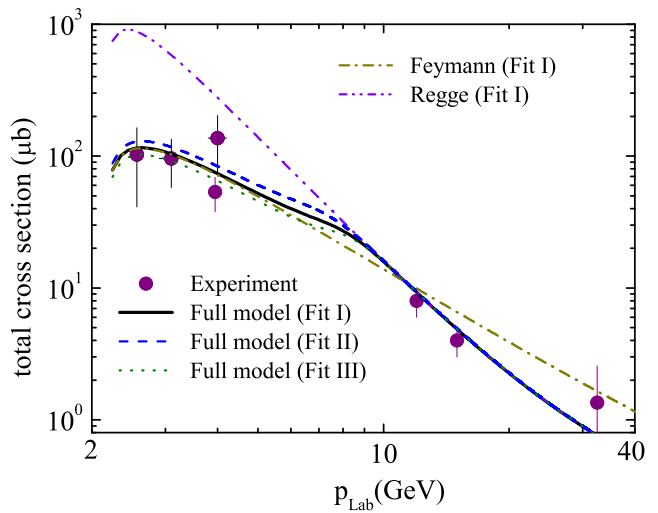}
\end{center}

\caption{Total cross section for the $\protect\pi ^{-}p\rightarrow f_{1}(1285)n$
reaction. The full (black), dashed (blue), dotted (green) dashed-dotted (dark
yellow), and dashed-dot-dotted (violet) lines are for the full model (Fit I),
full model (Fit II), full model (Fit III), Feynman case [$R(s,t)=0$] and Regge
case [$R(s,t)=1$], respectively. The experimental data are from Refs.~%
\protect\cite{Dahl:1967pg,Dionisi:1980hi,Bityukov:1983cw,Bityukov:1987bj}.}
\label{Fig:total1}
\end{figure}

In this subsection, we will fit all the data we collected as shown in Figs.~\ref{fig:data} and~\ref{Fig:total1},
which include four data points of the total cross section at low
beam momentum, three data points of the  total cross at high beam momentum, and four data
points of the $t$-distribution at 12-15 GeV~\cite%
{Dahl:1967pg,Corden:1978cz,Dionisi:1980hi,Bityukov:1983cw,Bityukov:1987bj}.
It should be mentioned that the three data points of the total cross section at
high beam momentum are obtained by continuation of the $t$-channel contribution at
very forward angles to all angles, so we will fit these three data points
only with the $t$-channel contribution because the $u$-channel contribution
is negligible at forward angles and dominant at backward angles. For the
three data points at low beam momentum, both $t$ and $u$ channels will be
included. It will be found later that the $s$-channel contribution is negligible,
as usual. We minimize $\chi ^{2}$ per degree of freedom by fitting five
parameters $s_{0}$, $s_{R}$, $\Lambda _{t}$, $\Lambda _{u}$ and $g_{a_{0}NN}$
using a total of 11 data points at the beam momentum $P_{Lab}$ from 2 to 40
GeV as displayed in Fig. \ref{Fig:total1}. Here,  $R(t)$ has been assumed to be  1 as discussed above.

\renewcommand\tabcolsep{0.54cm} \renewcommand{\arraystretch}{1.2}
\begin{table*}[htbp]
\caption{Fitted values of free parameters with all 11 data points in Refs.~%
\protect\cite%
{Dahl:1967pg,Corden:1978cz,Dionisi:1980hi,Bityukov:1983cw,Bityukov:1987bj}.}%
\label{Tab: parameterfull}
\begin{tabular}{|c|c|c|c|c|c|c|}
\hline\hline
& $\Lambda _{t}$ (GeV) & $\Lambda _{s}=\Lambda _{u}$ (GeV) & $s_{0}$ (GeV$%
^{2}$) & $s_{R}$ (GeV$^{2}$) & $g_{a_{0}NN}$ & $\chi ^{2}/d.o.f.$ \\ \hline
Fit I & $1.26\pm 0.02$ & $0.50\pm 0.78$ & $1.53\pm 0.55$ & $14.99\pm 1.47$ &
$28.44\pm 0.08$ & 1.21 \\
Fit II & $1.27\pm 0.01$ & $0.50\pm 0.77$ & $1.47\pm 0.38$ & $13.76\pm 1.38$
& $28.44\pm 0.11$ & 1.18 \\
Fit III & $1.25\pm 0.01$ & $0.50\pm 0.77$ & $1.35\pm 0.32$ & $15.46\pm 1.36$
& $28.44\pm 0.07$ & 1.16 \\ \hline
\end{tabular}%
\end{table*}

As observed in Fig.~\ref{Fig:total1}, at  $p_{Lab}=$ 3.95 and 4 GeV, there
exist two data points, which are quite different from each other. Because
the beam momenta of these two data points are very close, it is
difficult to interpret them as a physical structure. We present the results
by fitting with both data points at a beam  momentum of about 4 GeV (Fit I), the results
with a higher momentum (Fit II) and the results with a lower  momentum (Fit III). The
results suggest that the higher data point is difficult to  reproduce in
three fits, whose results are close to each other and support the lower data
point. The  fitted parameters are listed in Table~\ref{Tab: parameterfull}, and the values of the coupling constant $g_{a_0NN}$ and cutoff $\Lambda_t$ are close to those in Table~\ref{tab:fit}. We also present the results of the usual Feynman case [$R(s,t )=0$] and Regge
case [$R(s, t)=1$] in Fig. \ref{Fig:total1}, which show that the experimental data
of the total cross section of the $\pi ^{-}p\rightarrow f_{1}(1285)n$ reaction
cannot be reproduced using the Feynman model alone, even with the
traditional Reggeized treatment. The interpolating Reggeized treatment is
essential to reproduce the total cross section at both low and high beam momenta.

\begin{figure}[h!]
\begin{center}
\includegraphics[scale=0.305]{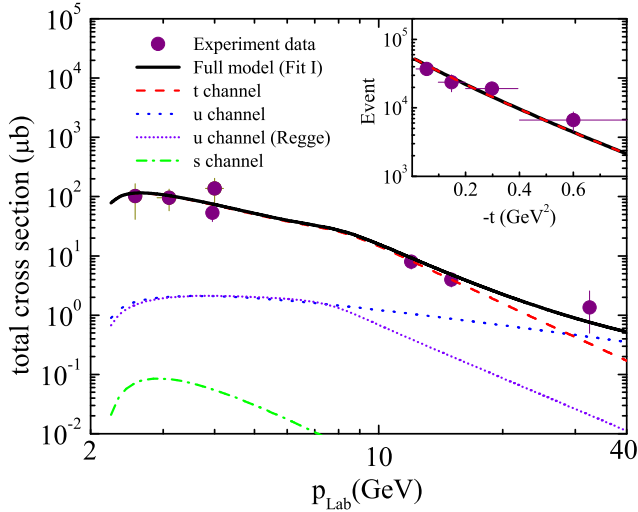}
\end{center}
\caption{Total cross section for the $\protect\pi ^{-}p\rightarrow f_{1}(1285)n$
reaction. The full (black), dashed (red), dotted (blue), short-dotted (violet), and dashed-dotted (green) lines are for
the full model (Fit I), $t$ channel, $u$ channel, $u$ channel with interpolated Reggeized treatment and $s$ channel, respectively. The
experimental data are from Refs.~\protect\cite%
{Dahl:1967pg,Corden:1978cz,Dionisi:1980hi,Bityukov:1983cw,Bityukov:1987bj}.}
\label{Fig:total2}
\end{figure}

In Fig.~\ref{Fig:total2}, we present the explicit results with Fit I. The
results show that the experimental data of both the total and $t$ distribution
can be well reproduced in our model. The $t$-channel contribution is
dominant at $p_{lab}$ up to about 20 GeV. The $u$-channel contribution is
negligible compared with the $t$-channel contribution at low beam momenta, but
becomes more important and exceeds the $t$-channel contribution at a beam momentum of
about 30 GeV.

\begin{figure}[h!]
\centering
\includegraphics[scale=0.305]{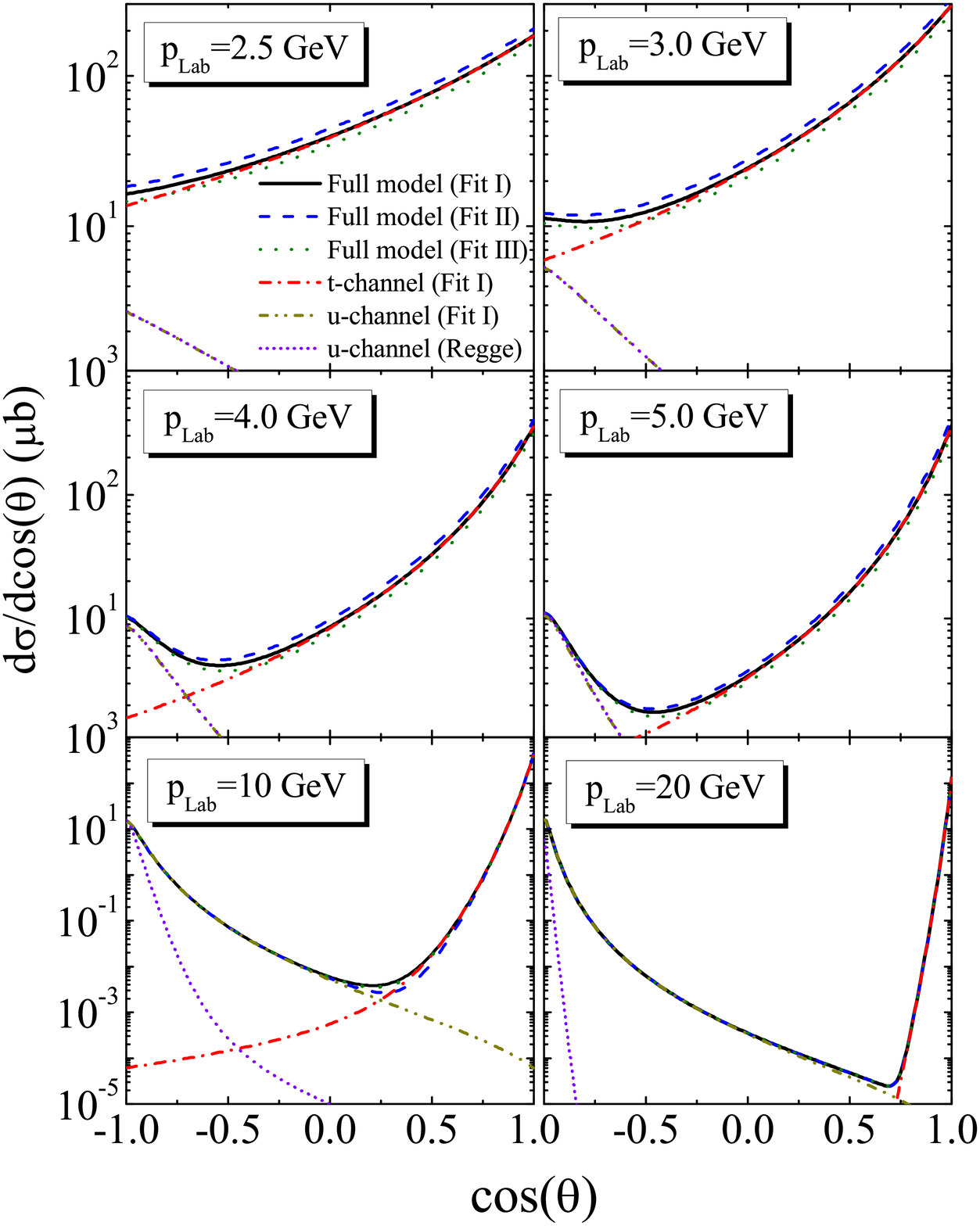}
\caption{The differential cross section $d\protect\sigma /d\cos \protect%
\theta $ of the $\protect\pi ^{-}p\rightarrow f_{1}(1285)n$ interaction as a
function of $\cos \protect\theta $. The full (black), dashed (blue), dotted
(green), dashed-dotted(red), dashed-dot-dotted (dark yellow), and short-dotted (violet)  lines are for
the full model (Fit I), full model (Fit II), full model (Fit III), $t$
channel (Fit I), $u$ channel (Fit I), and $u$ channel with  interpolated Reggeized treatment . }
\label{Fig: dcs}
\end{figure}

The $u$-channel contribution can be seen more clearly in the differential
cross section as shown in Fig.~\ref{Fig: dcs}.
The $t$ and $u$ channels
appear at forward and backward angles as expected. At low beam momentum, the
differential cross section is dominated by the $t$ channel at a large range of
the angles, whose contribution decreases with the decrease of the $\cos\theta
$. At momenta lower than about 3 GeV, the $t$ channel is more important
than the $u$ channel even at extreme backward angles. At the beam momenta higher
than about 3 GeV, the $u$ channel becomes more and more important with the
increase of the beam momentum. At a beam momentum of 20 GeV, though the total cross
section is still mainly from the $t$-channel contribution, the $u$ channel
is dominant even at medium angles while the $t$ channel is only dominant at
very forward angles. The results of the three fits are also presented in Fig.~%
\ref{Fig: dcs}. The discrepancy of the differential cross sections of three
fits is small at most beam momenta.

\subsection{Reggeized $u$-channel contribution}\label{Regu}
In the above calculation, the Reggeized treatment is applied to the $t$ channel, but not to the $u$ channel. Physically, the $u$ channel can be seen as a $t$ channel with the final particles interchanged.  Hence, the Reggeized treatment should be adopted in the $u$ channel, and as in the $t$ channel the interpolated treatment  is needed to connect the Regge case at high beam momentum smoothly to the Feynman case at low beam momentum~\cite{shk2015,bgy2011}.  Since the experimental data at high beam momenta are only obtained from the $t$-channel contribution, the fitting procedure in the above is not affected by the inclusion of the interpolated Reggeized  $u$-channel contribution, whose contribution at low beam momentum is just the Feynman type that we adopted in the above fitting procedure. However, the different treatment of the $u$ channel will affect prediction of the cross section at a high beam momentum, which will be discussed in this subsection.

The Reggeized treatment for $u$-channel baryon exchange
consists of replacing  the form factor $F_{u}(u)$ in Eq.~(\ref{AmpT2}) as
\begin{equation}
\mathcal{F}_{u}(u)=(\frac{s}{s_{scale}})^{\alpha
_{N}(u)-\frac{1}{2}}\frac{\pi \alpha _{N}^{\prime }(u-m_{N}^{2})}{\Gamma
\lbrack \frac{1}{2}+\alpha _{N}(u)]\sin [\pi (\alpha _{N}(u)-\frac{1}{2})]}.
\end{equation}%
 The scale factor $s_{scale}$ is fixed at 1 GeV. In addition, the Regge
trajectories $\alpha _{N}(u)$ read  \cite{PR1984}%
\begin{equation}
\alpha _{N}(u)=-0.34~{\rm GeV}^2 +0.99u.\quad \ \
\end{equation}

The interpolating can be applied to the $u$ channel analogously to the  $t$ channel  by replacing $t$  with $u$. It is reasonable to assume that the Reggeized treatment begins to exhibit its effect at the same value of beam momentum for both  the $t$ channel and $s$ channel. So, we adopt the same parameters in the interpolating treatment for the $t$ channel as those for the $s$ channel. As said above, the fitting procedure is not affected with the inclusion of the interpolated Reggeized treatment in the $u$ channel.  In this work, coupling constants involved in the $u$ channel are fixed  at the values  in our previous work~\cite{Wang:2017plf}. The $u$ channel at high beam momentum is determined after the cutoff $\Lambda_{u}$ is fixed in the fitting of the experimental data. In Fig.~\ref{Fig:total2},  the numerical results of the total cross section of the $u$ channel with interpolated Reggeized treatment are presented.  As expected, it decreases  exponentially as the $t$ channel at high beam momentum and is much smaller than those without Reggeized treatment.  However, the small contribution of the Reggeized $u$ channel does not mean that its contribution is negligible in the differential cross section. As shown in  Fig.~\ref{Fig: dcs}, the $u$ channel plays an important role in  shaping the differential cross section at backward angles at high beam momentum. The results in the full model are almost the sum of the $u$- and $t$-channel contributions, which are not given explicitly in the figures. 
Most of the $f(1285)$ events  are at extreme forward and backward angles, which correspond to the Reggeized $t$ and $u$ channel, respectively. At low beam momentum, the results with interpolated Reggeized treatment sre almost the same as those with  the Feynman type. 

\section{Summary and discussion}

In this work, based on the exisitng experimental data, we analyze the $\pi
^{-}p\rightarrow f_{1}(1285)n$ reaction with an interpolating Reggeized
approach and try to make a prediction of its total and differential cross
sections at a large beam-momentum range from threshold up to several tens of
GeV.
It is found that a pure Feynman or pure Regge type of  $t$-channel
contribution cannot reproduce the exisitng experiment data, though there are
only 11 data points against 5 free parameters. The interpolating
Reggeized treatment is essential to reproduce the cross sections at both low
and high beam momenta. 

At low momenta, both total and differential cross sections are
dominant with the Feynman-type $t$ channel. At
high beam momenta, the Reggeized $t$-channel contribution is only dominant at extreme forward angles and  decays rapidly  with the decease of  $\cos\theta$.  The $u$-channel contributions with and without Reggeized treatment exhibit  quite different behaviors at a high beam momentum.  Without the Reggeized treatment,   the $u$
channel becomes important in a larger range of angles with the increase of the beam momentum, while the $t$ channel
plays its role only at a very forward angle at high beam momenta. With the Reggeized treatment, the $u$ channel and $t$ channel provide a sharp increase and a sharp decrease at extreme backward and extreme forward angles, respectively. 

The low- and high-momentum pion beams are accessible at the J-PARC and COMPASS.
Our result is helpful to the possible experimental research of  $f_1(1285)
$ at  the two facilities. Based on the results, the measurement at forward
angles is supported while a measurement at extreme backward angles is helpful to understand the interaction mechanism of the pion-induced $f_1(1285)$ production. 

\section{Acknowledgments}

This project is supported by the National Natural Science Foundation of
China under Grant No. 11675228 and the Major State Basic Research Development
Program in China under Grant No. 2014CB845405.

\end{document}